\documentclass{Interspeech}

\interspeechcameraready

\title{EASY: Emotion-aware Speaker Anonymization via Factorized Distillation}

\author[affiliation={1}]{Jixun}{Yao}
\author[affiliation={2}]{Hexin}{Liu}
\author[affiliation={2}]{Eng Siong}{Chng}
\author[affiliation={1,*}]{Lei}{Xie}

\affiliation{Audio, Speech and Language Processing Group (ASLP@NPU)}{School of Computer Science}{Northwestern Polytechnical University, China}
\affiliation{}{Nanyang Technological University}{Singapore}

\email{yaojx@mail.nwpu.edu.cn, lxie@nwpu.edu.cn}


\usepackage{comment}
\usepackage{hyperref}
\usepackage{url}
\usepackage{algorithm}
\usepackage{algorithmic}
\usepackage{graphicx} 
\usepackage{float} 
\usepackage{subfigure} 
\usepackage{diagbox}
\usepackage{algorithm}
\usepackage{algorithmic}
\usepackage{multirow}
\usepackage{makecell}
\usepackage{amsmath}
\usepackage{caption}
\usepackage{cite}
\usepackage{color}
\usepackage{array}
\usepackage{amssymb}
\usepackage{hyperref}
\usepackage{booktabs}
\usepackage[table,xcdraw]{xcolor}
\usepackage{colortbl}
\usepackage{wrapfig}

\begin{document}

\maketitle

\begin{abstract}
    Emotion plays a significant role in speech interaction, conveyed through tone, pitch, and rhythm, enabling the expression of feelings and intentions beyond words to create a more personalized experience. However, most existing speaker anonymization systems employ parallel disentanglement methods, which only separate speech into linguistic content and speaker identity, often neglecting the preservation of the original emotional state. In this study, we introduce EASY, an emotion-aware speaker anonymization framework. EASY employs a novel sequential disentanglement process to disentangle speaker identity, linguistic content, and emotional representation, modeling each speech attribute in distinct subspaces through a factorized distillation approach. By independently constraining speaker identity and emotional representation, EASY minimizes information leakage, enhancing privacy protection while preserving original linguistic content and emotional state. Experimental results on the VoicePrivacy Challenge official datasets demonstrate that our proposed approach outperforms all baseline systems, effectively protecting speaker privacy while maintaining linguistic content and emotional state.
\end{abstract}

\section{Introduction}
With the advent of large language models, recent advancements in spoken dialogue systems, such as GPT-4o~\cite{gpt4o}, have demonstrated remarkable capabilities in human-computer interaction~\cite{defossez2024moshi,xie2024mini}. These models can perceive subtle emotional changes in speech and provide context-aware responses, significantly enhancing usability and interactivity across diverse real-world applications. However, these systems often require uploading speech data to cloud servers for processing, which poses substantial privacy risks, as such data can reveal sensitive personal information, including age, gender, health status, etc. Protecting sensitive information while preserving linguistic content and emotional state in the speech signal is essential for real-world speech interaction applications.

Existing privacy protection approaches for speech technology, known as speaker anonymization, aim to conceal the speaker's identity (privacy) while preserving linguistic content and other paralinguistic attributes unchanged (utility), such as emotion and gender~\cite{anon3_miao2024adapting}. The VoicePrivacy Challenge (VPC) series, introduced by the speech community and held over three editions~\cite{vpc2020,vpc2022,vpc2024}, strives to standardize speaker anonymization tasks, establish benchmarks, and encourage fair comparisons. Recent editions of the VPC have placed greater emphasis on not only preserving linguistic content but also retaining emotional states, highlighting the significance of emotion preservation in speaker anonymization.

Existing speaker anonymization approaches, as categorized by the VPC, fall into two main types: signal-processing-based approaches~\cite{sig1,sig2,sig3} and neural voice conversion-based approaches~\cite{nn1,nn2,yao2023distinguishable,anon1_meyer2023prosody,yao2024distinctive,anon2_panariello2024codec,yao2025musa,lv2023salt}. Signal-processing-based approaches usually do not require training data and directly modify instantaneous speech characteristics to achieve anonymization. However, they often suffer from content distortion and are ineffective against stronger attackers, as demonstrated in several studies~\cite{yao2022nwpu,yao2024npu_vpc}. In contrast, neural voice conversion-based approaches first disentangle speech into linguistic content and speaker identity, sometimes incorporating fundamental frequency modeling. The speaker identity is then replaced with an anonymized version to reconstruct the speech with the disentangled linguistic content. These approaches have shown significant advantages over their signal-processing counterparts in privacy protection and linguistic content preservation. However, most neural voice conversion-based speaker anonymization techniques neglect emotional modeling. While a recent study~\cite{anon3_miao2024adapting} explores emotional preservation in speaker anonymization and introduces an emotion compensation strategy as a post-processing step, the performance of emotion preservation in anonymized speech remains suboptimal.

To effectively conceal speaker identity while preserving the linguistic content and emotional state of the original speech, we propose an emotion-aware speaker anonymization framework called EASY. Unlike previous parallel disentanglement-based frameworks, our approach sequentially disentangles speaker identity, linguistic content, and emotional representation from the input waveform through a factorized distillation process, explicitly modeling each speech attribute in distinct subspaces. 
Specifically, we first disentangle speaker identity from the input speech signal through a supervised classifier functioning as an identity distillation mechanism, simultaneously producing a speaker-independent representation. Next, we leverage a residual vector quantization (RVQ) architecture, inherently suited for serial disentanglement, to disentangle linguistic content and emotional representation. For linguistic content distillation, we employ WavLM~\cite{chen2022wavlm} as the semantic teacher, while emotion2vec~\cite{ma2023emotion2vec} serves as the emotional teacher for emotion distillation. By independently constraining speaker identity and emotional representation, our framework minimizes information leakage between these attributes, thereby enhancing privacy protection while preserving linguistic and emotional integrity.
Experimental results on the VPC 2024 official evaluation datasets demonstrate that our proposed EASY framework outperforms all baseline systems in both privacy protection and emotion preservation. Additionally, ablation studies confirm the effectiveness of each component in the framework, highlighting their individual contributions to its overall performance.

\begin{figure*}[ht]
  \centering
  \includegraphics[width=14cm]{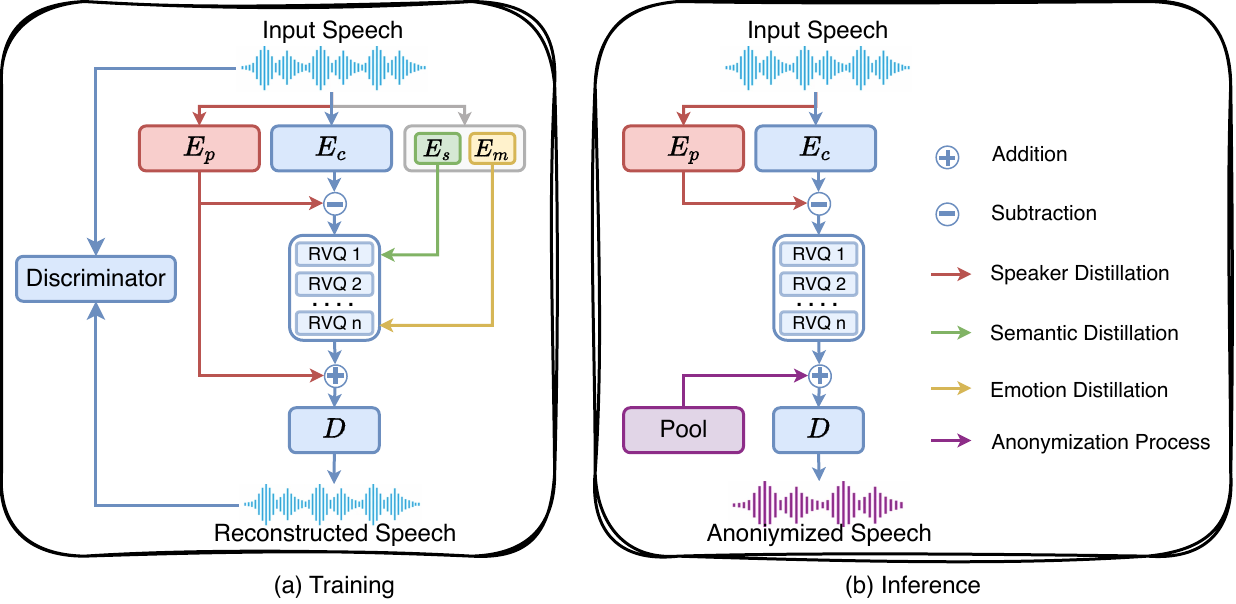}
  \caption{The overall architecture of our proposed speaker anonymization framework.}
  \label{fig:model}
  \vspace{-15pt}
\end{figure*}

\section{EASY}
\subsection{Overview}
Our proposed EASY framework employs an auto-encoder architecture comprising four individual encoders, a decoder, and a residual bottleneck module with multiple RVQ blocks. As illustrated in Figure~\ref{fig:model}, the speech encoder $E_c$ and speaker encoder $E_p$ are first employed to extract frame-level and utterance-level representations from the speech waveform. The frame-level representation captures various speech attributes, including linguistic content, speaker identity, and emotion. In contrast, the utterance-level representation mainly focuses on global attributes, such as speaker identity.


We serially disentangle each speech attribute to reconstruct the original waveform as follows: 1) we first constrain the utterance-level representation to speaker identity through supervised distillation, then subtract it from the frame-level representation to derive a speaker-independent representation; 2) we use a semantic encoder $E_s$ to distill linguistic content from the speaker-independent representation and employ a residual bottleneck module to capture residual information; 3) we distill the residual information with emotional representation using an emotion encoder $E_m$. Finally, the distilled attributes are concatenated and fed into the decoder $D$ to reconstruct the original speech waveform. The following sections will elaborate on the factorized distillation process, training objectives, and anonymization strategies.

\subsection{Factorized Distillation}
We introduce three distinct distillation methods to factorize speaker identity, linguistic content, and emotional representation from the input speech waveform.

\textbf{Speaker distillation}. Since speaker identity is an inherent, time-invariant characteristic of voice. We first transform the speech waveform into a mel-spectrogram and input it into our speaker encoder, which generates hidden sequences through fully connected layers, performing spectral modeling. We then use gated CNNs~\cite{gatedcnn} with residual connections to capture sequential information from these hidden sequences. Finally, several multi-head attention layers are applied, followed by an average pooling layer, to produce a global representation $s \in \mathbb{R}^{d}$, where $d$ represents the dimension of the speaker representation.

To ensure the extracted global representation primarily encodes speaker identity information while excluding other time-invariant attributes, we employ supervision distillation losses to constrain the output of the speaker encoder. A speaker classifier $C$ is introduced to align the global representation with explicit speaker identity labels $I$. The supervision distillation losses $\mathcal{L}_{\textrm{spk}}$ for speaker identity are defined as follows:
\begin{equation}
    \mathcal{L}_{\textrm{spk}}=\mathbb{E}[-log(C(I\mid s))].
\end{equation}
The speaker classifier determines whether $s$ is associated with the corresponding speaker identity. We then obtain the speaker identity representation and directly subtract it from the output of the speech encoder to obtain the speaker-independent representation $r_1$ for the next step of factorized distillation.

\textbf{Linguistic distillation}.
The residual bottleneck module consists of $N$ Vector Quantization (VQ) layers~\cite{van2017neural}, each cascading in a residual manner. Each quantizer contains only residual information from the preceding quantizer, making it inherently suitable for serial disentanglement. We leverage this characteristic for hierarchical disentanglement to further separate linguistic content and emotional state from $r_1$. Following \cite{zhang2023speechtokenizer}, we choose the first VQ layer to represent linguistic content and introduce semantic distillation using the semantic encoder $E_s$. We use WavLM~\cite{chen2022wavlm} as our semantic encoder and extract the output from the 6th layer. We then apply k-means clustering to obtain semantic tokens, which serve as the distillation label. The linguistic distillation loss $\mathcal{L}_{\textrm{lin}}$ can be described as follows:
\begin{equation}
    \mathcal{L}_{\textrm{lin}}=\mathbb{E}[-log(\text{km}(w)\mid q_1))],
\end{equation}
where $\text{km}(\cdot)$ denote as K-means cluster, $q_1$ and $w$ represent the quantized token from the first quantizer and the output of 6th WavLM's layer, respectively.

\textbf{Emotion distillation}.
After aligning the first quantizer to linguistic content, the residual information contains minimal speaker identity and linguistic content. We introduce emotional distillation to constrain the remaining quantizers, ensuring they explicitly capture emotional representation while reducing leakage of speaker identity or linguistic content. We use emotion2vec as the emotional encoder, which acts as the distillation teacher for the residual quantizer. Fine-grained emotional representations are extracted from the emotional encoder $E_m$, and KL-divergence is applied to align the quantized embedding from the residual quantizer with these representations. To further prevent information leakage of linguistic content or speaker identity in residual quantizers, we employ an adversarial classifier with a gradient reversal layer (GRL) to enhance distillation performance. Specifically, a speaker-GRL predicts speaker identity labels to eliminate speaker identity information, while a semantic-GRL predicts semantic tokens to reduce linguistic content information. The emotional distillation loss $\mathcal{L}_{\textrm{emo}}$ is summarized as follows:
\begin{align}
    \mathcal{L}_{\textrm{emo}}= \text{KL}&(e,q_n) + \mathbb{E}_{\text{GRL}}[-log(\text{km}(w)\mid q_n)] \notag \\&+\mathbb{E}_{\text{GRL}}[-log(C(I\mid q_n))],
\end{align}
where $\text{KL}$ and $q_n$ represent KL-divergence and the quantized embedding of $n$-th quantizer.

\subsection{Training Objectives}
The training objective of EASY includes a reconstruction task with adversarial training and three factorized distillation losses. The reconstruction loss combines L1 and L2 distances in the frequency domain, which can be described as follows:
\begin{align}
    \mathcal{L}_{\textrm{rec}}=\Vert\text{mel}(X)-\text{mel}(\hat{X})\Vert_1  
    +\Vert\text{mel}(X)-\text{mel}(\hat{X})\Vert_2,
\end{align}
where $X$ and $\hat{X}$ represent the input speech waveform and reconstructed speech waveform, respectively. The function $\text{mel}(\cdot)$ represents an 80-bin mel-spectrogram extraction process, performed using a short-time Fourier transform (STFT).

To improve the quality of the reconstructed speech waveform, we incorporate an adversarial loss $\mathcal{L}_{\textrm{adv}}$, adopting the same formulation as in HiFi-GAN~\cite{hifigan}. Additionally, we use the straight-through estimator to optimize the commitment loss $\mathcal{L}_{\textrm{com}}$ between the input feature and the quantized feature in the residual bottleneck module as follows:
\begin{equation}
    \mathcal{L}_{\textrm{com}}=\sum_{i=1}^{N}||x_i-q_i||_2^2,
\end{equation}
where $i$ represents the number of quantizer layers.
Finally, the proposed framework is optimized using the weighted sum of the following losses:
\begin{equation}
\mathcal{L}=\lambda_r\mathcal{L}_{\textrm{rec}}
    +\mathcal{L}_{\textrm{adv}}
    +\mathcal{L}_{\textrm{com}} 
    +\mathcal{L}_{\textrm{spk}}
    +\mathcal{L}_{\textrm{lin}}
    +\mathcal{L}_{\textrm{emo}},
\end{equation}
where $\lambda_r$ is hyper-parameters used to balance the loss term, we set $\lambda_r=45$ following the previous study~\cite{yang2023hifi}. 

\subsection{Anonymization Process}
During inference, the semantic and emotion encoders are not required, as illustrated in Figure~\ref{fig:model}(b). Instead, an additional speaker vector pool containing various speaker identity representations extracted from different speakers is introduced.
Our proposed anonymization strategy involves a weighted sum of two components: the averaged speaker identity and a randomly generated speaker identity. To generate the averaged speaker identity, a set of speaker identities is randomly selected from the speaker vector pool and averaged to create a pseudo speaker identity that does not correspond to any real speaker. For the randomly generated speaker identity, we directly sample from a Gaussian distribution. The anonymization process can be described as follows:
\begin{equation}
    s_{\text{anon}} = \alpha\Bar{s}+(1-\alpha)\hat{s},
\end{equation}
where $\alpha$ denotes the weight parameter, $\Bar{s}$ and $\hat{s}$ represent the averaged and sampled speaker identities, respectively. 

\section{Experimental Setup}
\subsection{Dataset}
We train EASY using the training portions of both the LibriSpeech~\cite{panayotov2015librispeech} and LibriTTS~\cite{zen2019libritts} corpora. For evaluation, we follow the VPC 2024 configuration~\cite{vpc2024}, using LibriSpeech-dev-clean and LibriSpeech-test-clean for privacy and utility evaluation. Additionally, the IEMOCAP~\cite{busso2008iemocap} development and evaluation sets are used for emotion-related evaluation.

For model training, we use the AdamW optimizer with parameters $\beta_1$ = 0.8, $\beta_2$ = 0.99, and weight decay $\lambda$ = 1e-5. The learning rate follows a decay schedule, decreasing by a factor of 0.99 per epoch, starting from an initial learning rate of $2 \times 10^{-4}$. 
The training process comprised 50k steps, utilizing 4 NVIDIA 3090 GPUs with a batch size of 64 utterances. Additionally, we employ 8 quantizers in the residual bottleneck module.

\subsection{Evaluation Metrics}
Following the same settings as VPC 2024, we use three evaluation metrics: equal error rate (EER) for privacy evaluation, word error rate (WER) for linguistic content preservation, and unweighted average recall (UAR) for emotional preservation. A higher EER indicates stronger privacy protection, a lower WER reflects better linguistic content preservation, and a higher UAR signifies improved emotion preservation.

\subsection{Baseline Systems}
To comprehensively evaluate the performance of our proposed EASY, we compare it with several baseline systems. \textit{Official baseline systems}: All official baseline systems~\cite{vpc2020,sig1,anon1_meyer2023prosody,anon2_panariello2024codec,b5b6} from the VPC 2024 are included in our experiments, denoted as \textbf{B*}. \textit{Participant systems}: The ranked top three anonymization systems~\cite{vpc_t8,vpc_t9,yao2024npu_vpc} from VPC 2024 are selected as strong baselines, denoted as \textbf{T*}. \textit{Recently emotional anonymization system}: EASY is also compared to the most recent emotional speaker anonymization system~\cite{anon3_miao2024adapting}, denoted as \textbf{OH}.

\begin{table*}[ht]
\centering
\renewcommand\arraystretch{1.2}
\caption{Averaged results between baseline systems and EASY on the VPC development and test datasets. For EER and UAR, higher scores indicate better performance, while for WER, lower scores are preferable.}
\label{tab:comp}
\begin{tabular}{lcccccc}
\hline
      & \multicolumn{2}{c}{EER, \% ($\uparrow$)}            & \multicolumn{2}{c}{WER, \% ($\downarrow$)}            & \multicolumn{2}{c}{UAR, \% ($\uparrow$)}    \\ \cline{2-7} 
      & LibriSpeech-dev & LibriSpeech-test & LibriSpeech-dev & LibriSpeech-test & IEMOCAP-dev & IEMOCAP-test \\ \hline
Orig. & 5.72            & 4.59             & 1.80            & 1.85             & 69.08       & 71.06        \\ \hline
B1~\cite{vpc2020}    & 9.20\textcolor{gray}{$_{+0.0}$} & 6.07\textcolor{gray}{$_{+0.0}$} & 3.07\textcolor{gray}{$_{+0.0}$} & 2.91\textcolor{gray}{$_{+0.0}$} & 42.71\textcolor{gray}{$_{+0.0}$} & 42.78\textcolor{gray}{$_{+0.0}$} \\
B2~\cite{sig1}    & 7.48\textcolor{teal}{$_{-1.72}$} & 4.52\textcolor{teal}{$_{-1.55}$} & 10.44\textcolor{purple}{$_{+7.37}$} & 9.95\textcolor{purple}{$_{+7.04}$} & 55.61\textcolor{purple}{$_{+12.9}$} & 53.49\textcolor{purple}{$_{+10.71}$} \\
B3~\cite{anon1_meyer2023prosody}    & 25.24\textcolor{purple}{$_{+16.04}$} & 27.32\textcolor{purple}{$_{+21.25}$} & 4.29\textcolor{purple}{$_{+1.22}$} & 4.35\textcolor{purple}{$_{+1.44}$} & 38.09\textcolor{teal}{$_{-4.62}$} & 37.57\textcolor{teal}{$_{-5.21}$} \\
B4~\cite{anon2_panariello2024codec}    & 32.71\textcolor{purple}{$_{+23.51}$} & 30.26\textcolor{purple}{$_{+24.19}$} & 6.15\textcolor{purple}{$_{+3.08}$} & 5.90\textcolor{purple}{$_{+2.99}$} & 41.97\textcolor{teal}{$_{-0.74}$} & 42.78\textcolor{gray}{$_{+0.0}$} \\
B5~\cite{b5b6}    & 34.37\textcolor{purple}{$_{+25.17}$} & 34.34\textcolor{purple}{$_{+28.27}$} & 4.73\textcolor{purple}{$_{+1.66}$} & 4.37\textcolor{purple}{$_{+1.46}$} & 38.08\textcolor{teal}{$_{-4.63}$} & 38.17\textcolor{teal}{$_{-4.61}$} \\
B6~\cite{b5b6}    & 23.05\textcolor{purple}{$_{+13.85}$} & 21.14\textcolor{purple}{$_{+15.07}$} & 9.69\textcolor{purple}{$_{+6.62}$} & 9.09\textcolor{purple}{$_{+6.18}$} & 36.39\textcolor{teal}{$_{-6.32}$} & 36.13\textcolor{teal}{$_{-6.65}$} \\ \hline
T8~\cite{vpc_t8}    & 40.93\textcolor{purple}{$_{+31.73}$} & 40.70\textcolor{purple}{$_{+34.63}$} & 3.45\textcolor{purple}{$_{+0.38}$} & 3.19\textcolor{purple}{$_{+0.28}$} & 47.07\textcolor{purple}{$_{+4.36}$} & 47.10\textcolor{purple}{$_{+4.32}$} \\
T9~\cite{vpc_t9}    & 33.43\textcolor{purple}{$_{+24.23}$} & 35.10\textcolor{purple}{$_{+29.03}$} & \textbf{2.33\textcolor{teal}{$_{-0.74}$}} & \textbf{2.37\textcolor{teal}{$_{-0.54}$}} & 60.69\textcolor{purple}{$_{+17.98}$} & 60.95\textcolor{purple}{$_{+18.17}$} \\
T10~\cite{yao2024npu_vpc}   & 42.45\textcolor{purple}{$_{+33.25}$} & 40.46\textcolor{purple}{$_{+34.39}$} & 3.51\textcolor{purple}{$_{+0.44}$} & 3.19\textcolor{purple}{$_{+0.28}$} & 62.93\textcolor{purple}{$_{+20.22}$} & 60.87\textcolor{purple}{$_{+18.09}$} \\
OH~\cite{anon3_miao2024adapting}   & 39.92\textcolor{purple}{$_{+30.72}$} & 38.54\textcolor{purple}{$_{+32.47}$} &  2.36\textcolor{teal}{$_{-0.71}$} &  2.48\textcolor{teal}{$_{-0.43}$} & 47.01\textcolor{purple}{$_{+4.40}$} & 47.37\textcolor{purple}{$_{+4.59}$} \\ \hline
EASY  & \textbf{46.67\textcolor{purple}{$_{+37.47}$}} & \textbf{45.10\textcolor{purple}{$_{+39.03}$}} & 2.71\textcolor{teal}{$_{-0.36}$} & 2.69\textcolor{teal}{$_{-0.22}$} & \textbf{64.19\textcolor{purple}{$_{+21.48}$}} & \textbf{63.43\textcolor{purple}{$_{+20.65}$}} \\ \hline
\end{tabular}
\vspace{-10pt}
\end{table*}

\section{Experimental Results}
\subsection{Privacy and Utility Evaluation}

We evaluate EASY using the VPC evaluation datasets and compare its performance with the official VPC baseline systems and state-of-the-art anonymization systems. 
As shown in Table~\ref{tab:comp}, the EER results for B1 and B2 are only slightly higher than those of the original speech, indicating poor privacy protection performance. In contrast, EASY achieves EERs of 46.67\% and 45.10\% on the test datasets, significantly outperforming all baseline systems. These results demonstrate that the speaker verification model is unable to recognize the original speaker identity in speech anonymized by EASY, highlighting the effectiveness of our approach in concealing speaker identity and protecting personal privacy.

In terms of linguistic content preservation, EASY achieves superior WER results compared to all VPC official baseline systems, with values only slightly higher than those of the original speech. These results demonstrate that EASY effectively preserves linguistic content, achieving better utility performance. Regarding emotional preservation, EASY achieves the highest UAR results among all baseline systems, with scores of 64.19\% and 63.43\% on the development and test sets, respectively. Although the OH system achieves slightly lower WER results than EASY, its performance in emotional preservation is worse than our proposed systems. These findings indicate that EASY outperforms baseline systems in utility preservation, successfully maintaining both linguistic content and emotional states.

\subsection{Ablation Studies}
We investigate the effectiveness of the three specially designed factorized distillation methods for insights into their contributions to the overall performance. As shown in Table~\ref{tab:ablation}, the results reveal the following: 1) removing speaker identity distillation $\mathcal{L}_{\textrm{spk}}$ leads to degraded EER and UAR results, indicating that the output of the speaker encoder retains both speaker identity and emotion information, causing some original emotional information to be lost during the anonymization process; 2) removing linguistic content distillation $\mathcal{L}_{\textrm{lin}}$ significantly worsens WER results, reflecting reduced intelligibility of the anonymized speech. This distortion also slightly increases EER results; 3) removing emotional distillation $\mathcal{L}_{\textrm{emo}}$ results in lower EER and UAR scores, demonstrating that employing both speaker and emotion distillation achieves better disentanglement and ensures minimal information leakage.
These results highlight the importance of factorized distillation in achieving superior disentanglement performance in our anonymization framework.

\begin{table}[h]
\centering
\renewcommand\arraystretch{1.1}
\caption{Results of ablation studies on different factorized distillation variants.}\label{tab:ablation}
\begin{tabular}{lccc}
\hline
     & EER, \% ($\uparrow$)   & WER, \% ($\downarrow$)  & UAR, \% ($\uparrow$)   \\ \hline
EASY & 45.89 & 2.70 & 63.81 \\
$\quad$ w/o $\mathcal{L}_{\textrm{spk}}$  & 31.84  & 2.97 & 45.36  \\
$\quad$ w/o $\mathcal{L}_{\textrm{lin}}$  & 46.83  & 6.21 & 43.57  \\
$\quad$ w/o $\mathcal{L}_{\textrm{emo}}$  & 42.38  & 2.84 & 60.05  \\ \hline
\end{tabular}
\end{table}

We further analyze the information preserved in each VQ layer by reconstructing the input waveform using the particular VQ layer. As shown in Table~\ref{tab:vq}, reconstructed speech using VQ-1 achieves lower WER compared to VQ-2:8, indicating that the first VQ layer primarily preserves linguistic information. Conversely, the UAR for VQ-2:8 outperforms VQ-1, suggesting that the residual VQ layers contain more emotion-related information. VQ-1 and VQ-2:8 both show significantly degraded performance compared to VQ-1:8, which effectively preserves both linguistic and emotional content. This degradation occurs because VQ-1 lacks emotional information, while VQ-2:8 lacks linguistic content.
When speaker identity representation is removed (w/o $s$), the EER values increase significantly, demonstrating that the speaker encoder effectively captures speaker identity information.

\begin{table}[h]
\centering
\renewcommand\arraystretch{1.1}
\caption{Ablation studies on the disentangling performance of each VQ layer.}\label{tab:vq}
\begin{tabular}{lccc}
\hline
     & EER, \% ($\uparrow$)   & WER, \% ($\downarrow$)  & UAR, \% ($\uparrow$)   \\ \hline
VQ-1:8 & 5.16 & 1.83 & 70.07 \\
VQ-1  & 14.68  & 13.48 & 32.79  \\
VQ-2:8  & 24.72  & 26.54 & 41.28  \\ 
w/o $s$  & 29.46  & 2.72 & 63.11  \\ \hline
\end{tabular}
\vspace{-10pt}
\end{table}
\section{Conclusion}
In this study, we present EASY, an emotion-aware speaker anonymization framework that sequentially disentangles speaker identity, linguistic content, and emotional representation through a factorized distillation process, explicitly modeling each speech attribute in distinct subspaces. Speaker identity is first disentangled using a supervised classifier, producing a speaker-independent representation. For linguistic content and emotion, we leverage two pre-trained self-supervised models as distillation teachers to separate these attributes.
By independently constraining each speech attribute, EASY minimizes information leakage, achieving enhanced privacy protection while preserving linguistic content and emotional state. Experimental results on the VoicePrivacy 2024 official evaluation datasets demonstrate that EASY surpasses all baseline systems in privacy protection and emotion preservation. Furthermore, ablation studies validate the effectiveness of each component in the framework.

\section{Acknowledgement}
This work was supported by Tencent and Tencent-NTU Joint Research Laboratory (CENTURY), Nanyang Technological University, Singapore.

\bibliographystyle{IEEEtran}
\bibliography{mybib}

\end{document}